\documentclass[12pt,preprint]{aastex}

\begin{document}

\title{$^{128}$Xe and $^{130}$Xe: Testing He-shell burning in AGB stars}

\author{R. REIFARTH}
\affil{Los Alamos National Laboratory, Los Alamos, New Mexico, 87545, USA}
\email{reifarth@lanl.gov}

\author{F. K\"{A}PPELER, F. VOSS, K. WISSHAK}
\affil{Forschungszentrum Karlsruhe, Institut f\"{u}r Kernphysik,
       Postfach 3640, D-76021 Karlsruhe, Germany}

\author{R. GALLINO, M. PIGNATARI}
\affil{Dipartimento di Fisica Generale, Universit{\`a} di Torino, Sezione
INFN di Torino, Via P. Giuria 1, I-10125 
       Torino, Italy and Centre for Stellar and Planetary Astrophysics, 
	   School of Mathematical Sciences, Monash University 3800, Victoria, Australia}

\author{O. STRANIERO}
\affil{Osservatorio Astronomico di Collurania, I-64100 Teramo, Italy}

\begin{abstract}
The $s$-process branching at $^{128}$I has been investigated on the basis of new, 
precise experimental ($n, \gamma$) cross sections for the $s$-only isotopes $^{128}$Xe
and $^{130}$Xe. This branching is unique, since it is essentially determined  
by the temperature- and density-sensitive stellar decay rates of $^{128}$I and  
only marginally affected by the specific stellar neutron flux. For this 
reason it represents an important test for He-shell burning in 
AGB stars. The description of the branching by means of the complex stellar scenario
reveals a significant sensitivity to the time scales for convection during
He shell flashes, thus providing constraints for this phenomenon. 
The $s$-process ratio $^{128}$Xe/$^{130}$Xe deduced from stellar models allows 
for a 9$\pm$3\% $p$-process contribution to solar $^{128}$Xe, in agreement with 
the Xe-S component found in meteoritic presolar SiC grains.
\end{abstract}


\section{Introduction}

\begin{figure}
\plotone{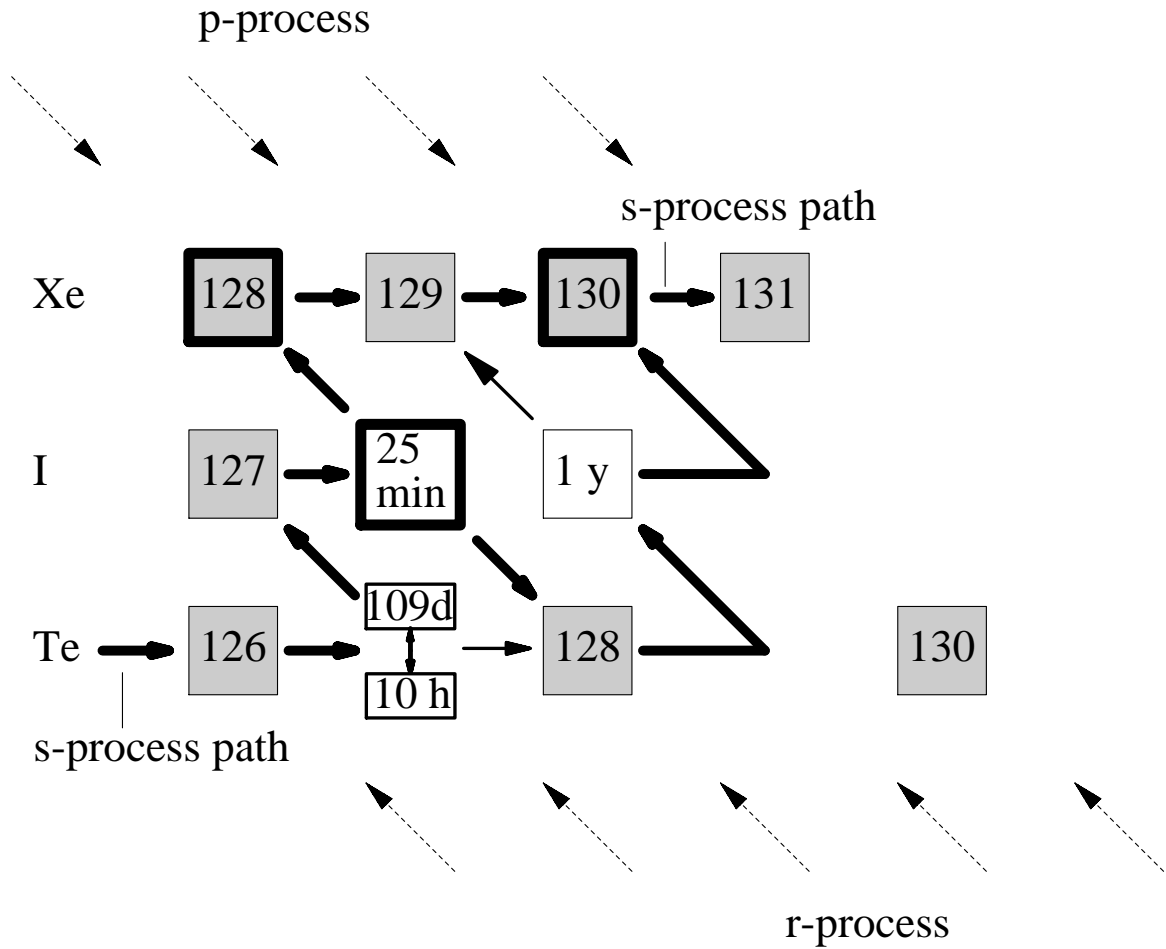}
\caption{The $s$-process reaction path between Te and Xe. The isotopes 
  $^{128}$Xe and $^{130}$Xe are shielded against $r$-process contributions by 
  their stable Te isobars. In contrast to $^{130}$Xe, $^{128}$Xe is partly 
  bypassed due to the branching at $^{128}$I. The branching at $^{127}$Te is 
  negligible unless the temperature is low enough that ground state and isomer 
  are not fully thermalized. The branching at $^{128}$I is unique since it 
  results from the competition between ${\beta^-}$ and electron capture decays,
  independent of the neutron flux. 
\label{fig1}}
\end{figure}

The long chain of the stable Xe isotopes exhibits the signatures of all
scenarios contributing to the production of heavy elements beyond Fe, and 
is, therefore, of highest astrophysical interest. The light isotopes, 
$^{124}$Xe and $^{126}$Xe can be assigned to the $p$ process, since they
can not be produced via neutron capture. Their relative isotopic abundances 
are important for testing nucleosynthesis models describing the proton-rich 
side of the stability valley. Concerning the $s$ process, xenon belongs to 
the six elements with a pair of $s$-only isotopes. In this case, the relevant 
nuclei are $^{128}$Xe and $^{130}$Xe, both shielded against the $r$-process 
region by their stable Te isobars. The abundances of these 
isotopes define the strength of the branching in the $s$-process reaction chain 
illustrated in Fig.\ref{fig1}. Since the $p$-process contribution to the observed abundances 
in this mass region can typically be neglected, such shielded nuclei are commonly considered 
to be of pure $s$ origin. In the case of Xe, however, it was pointed out that 
$^{128}$Xe is not a typical $s$-only nucleus, since its solar abundance may
include a significant $p$ contribution \cite{WoH78}. On the neutron-rich side, 
$^{134}$Xe and $^{136}$Xe are considered as $r$-only nuclei, since the 
$\beta^-$ half life of $^{133}$Xe is short enough to prevent any significant 
$s$-process production. 

This paper aims
at a thorough discussion of the $s$-process aspects by using accurate
stellar ($n, \gamma$) cross sections of the relevant Te and Xe isotopes \cite{ReK02,RHK02}. 
Following a summary of the various sources of isotopic Xe abundance patterns 
($\S$\ref{facts}), the particular features of the $s$-process branchings at $A=127/128$
are discussed in $\S$\ref{sprocess}. The $s$-process aspects related to thermally 
pulsing asymptotic giant branch (AGB) stars are described in $\S$\ref{models},
and the results are presented in $\S$\ref{results}. 

\section{Facts and observations \label{facts}}

When stellar ($n, \gamma$) cross sections for $^{128, 130}$Xe with relative uncertainties 
of less then 2\% became available \cite{RHK02},
the determination of the elemental solar xenon abundance was a first natural application. 
While it is impossible to use standard methods for noble gases, i.e. meteorite analyses or solar 
spectroscopy, the solar Xe abundance of 5.30 $\pm$ 0.14 relative to Si=10$^6$ could be 
derived from $s$-process systematics. Apart from the solar abundance, these cross 
sections provide the key for the interpretation of the xenon isotope patterns, which bear 
promising clues for the analysis of presolar grains.

The noble gas nature of xenon implies that there are no stable chemical 
compounds. Accordingly, the isotopic ratios of xenon in planetary bodies were 
subject to mass fractionation effects during the earliest stages of the 
solar system. Terrestrial ratios are additionally affected by 
fission of U and Th, which are both enriched compared to 
the solar average \cite{DeB91, AnG89}. So far the most representative ratio
for the solar system,
$^{128}$Xe/$^{130}$Xe = 0.510$\pm$0.005, has been obtained from the solar wind 
component implanted in lunar rocks (Pepin et al. 1995; U. Ott, private communication). 
In the near future, more data are to be expected from the GENESIS mission \cite{RNN96}.  

\begin{figure}
\plotone{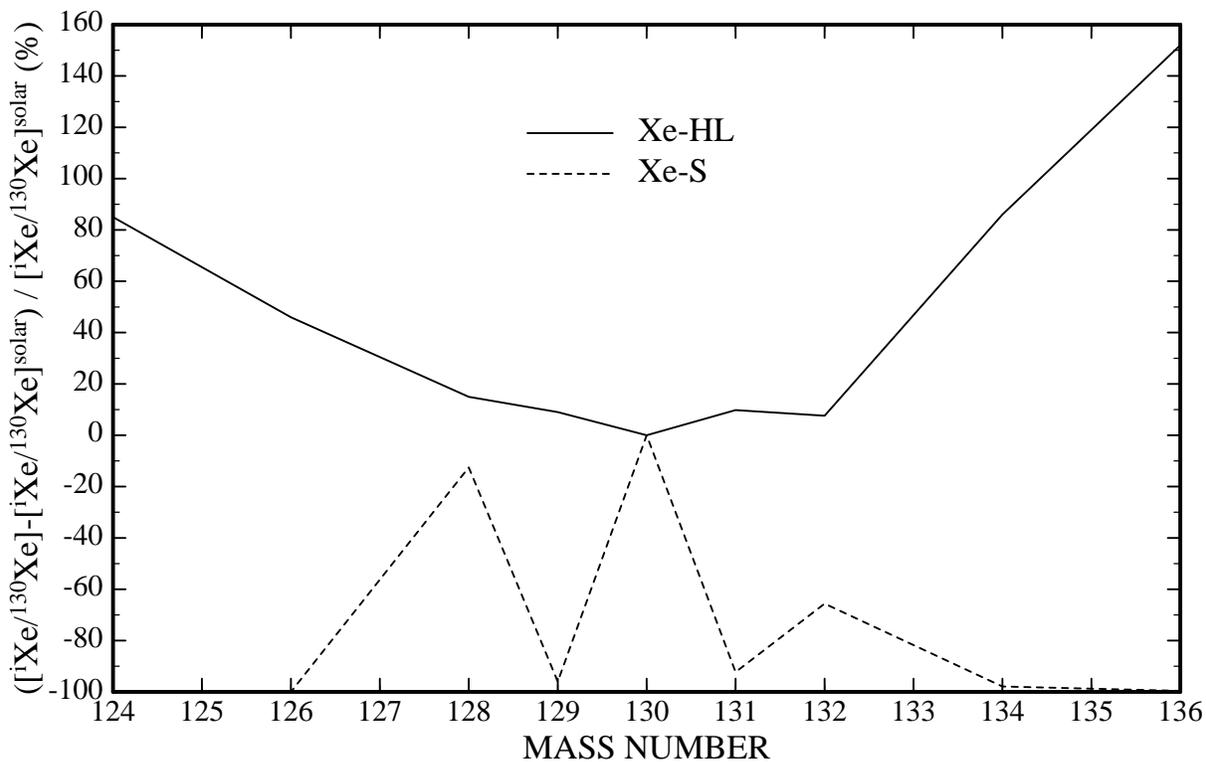}
\caption{The abundance pattern of Xe-S \protect\cite{LAA94} relative to the solar 
distribution shows that the ratio of the $s$ process isotopes 128 and 130 is 12\%
less than unity, indicating the effect of the $^{128}$I branching. The $s$ 
contributions to the other Xe isotopes are much smaller. Xe-HL, the counterpart from 
explosive nucleosynthesis is enhanced in the light and heavy isotopes produced in 
the $p$ and $r$ process, respectively \protect\cite{HuL94}. Fission
may additionally contribute to $^{129,131-136}$Xe. The abundance patterns are 
normalized at $^{130}$Xe. \label{fig2}}
\end{figure}

The discovery that presolar grains are carriers of noble gases with isotopic
compositions significantly different from solar material (Ott 1993; Zinner 
1998, and references therein) opened a new window to stellar and galactic
evolution. These grains, which originate from circumstellar
envelopes of AGB stars or supernova outflows by condensation of the most 
stable compounds, such as silicon carbide (SiC) and diamond (C), acted as 
carriers of trace elements like noble gases. Accordingly, their isotopic
composition contains a wealth of information on the nucleosynthesis at the site  
of their origin. Among the many identified elements in these grains, the isotope 
patterns of the noble gases are particularly prominent. Two 
characteristic components were isolated for xenon (see Fig.~\ref{fig2} and Table \ref{tab1}):       

- Xenon-S \cite{ReT64} was found to be carried by presolar mainstream
SiC grains 
(Lewis et al. 1994 and references therein), exhibits a pronounced 
zig-zag abundance pattern, reflecting the inverse of the respective 
stellar ($n, \gamma$) cross sections. This is characteristic for an $s$-process 
origin and points to AGB stars as the production site. 

- Xenon-HL, which was identified in presolar diamond grains 
(Huss \& Lewis 1994 and references therein). It exhibits enhanced 
abundances of the light and heavy isotopes and is most likely produced in 
supernova explosions, which are commonly believed to produce $r$- and $p$-enriched material.

Nucleosynthesis models are challenged by these 
observations, because the information contained in this material
witnesses the original on-site production. As far as the $s$-process 
component is concerned, the challenge is directly linked to the reliability
of the stellar neutron capture cross sections. The case of xenon is appealing, 
since the abundance ratio of the 
$s$-only nuclei $^{128}$Xe and $^{130}$Xe reflects the strength of 
the $^{127}$Te and $^{128}$I branchings as sketched in Fig.~\ref{fig1}. 

The presently available information on the $^{128}$Xe/$^{130}$Xe ratio  
for Xe-S (Table \ref{tab1}) can not be explained by the schematic 
classical approach of the $s$ process, which unavoidably overestimates this 
ratio at the characteristic temperature derived from the analysis of other 
branchings. More specific stellar evolutionary 
models have been invoked, using the Xe abundance patterns as a crucial 
model test.       

\section{The $s$-process branchings at $^{127}$Te and $^{128}$I \label{sprocess}}

As illustrated in Fig.~\ref{fig1}, the branchings at $^{127}$Te and $^{128}$I are expected to 
be comparably weak since only a small part of the total $s$-process flow is 
bypassing $^{128}$Xe. Therefore, the product of the 
stellar (n,$\gamma$) cross section and the respective $s$ abundance, 
which characterizes the reaction flow, is slightly smaller for $^{128}$Xe than 
for $^{130}$Xe. Since their solar isotopic abundance ratio \cite{PBR95} as well as the
stellar cross section ratio are accurately known, the $\langle\sigma\rangle N_s $ 
ratio, and hence the strength of the branching, 
can be determined. 

The branching at $^{127}$Te is weak, because the population of 
ground state and isomer is quickly thermalized in the hot
stellar photon bath, leading to a strong dominance of the $\beta$-decay
channel of the fast ground state decay. The neutron capture cross
section of $^{127}$Te is only theoretically known and may 
be uncertain by a factor of two. This branching plays an almost negligible 
role for the final $^{128}$Xe/$^{130}$Xe ratio as we shall discuss later.

The second branching at $^{128}$I with an half-life of only 25~min is 
exceptional, since it
originates from the competition between the short-lived $\beta^-$ and 
electron capture (EC) decays only. In contrast to 
other branchings, the influence of the stellar neutron flux is 
negligible in this case, thus eliminating an important uncertainty
in the $s$-process calculation of the isotopic Xe abundances. This 
provides a unique possibility to better constrain temperature and electron
density of the stellar plasma, which are manifested via the EC rate of $^{128}$I 
\cite{TaY87}.

The branching factor at $^{128}$I is
$$    
f_- = \frac{\lambda_{\beta^{-}}} {\lambda_{\beta^{-}}+\lambda_{\beta^{EC}}} = 
1- \frac{\lambda_{\beta^{EC}}} {\lambda_{\beta^{-}}+\lambda_{\beta^{EC}}}.$$
The $\beta^-$-decay rate varies only weakly with stellar temperature, but 
the electron capture rate depends 
strongly on temperature due to the increasing degree of ionisation. Furthermore,
at high temperatures, when the ions are fully stripped, the EC rate becomes 
sensitive to the density in the stellar plasma due to electron capture from the 
continuum. According to Takahashi \& Yokoi (1987), the EC rate decreases by 
one order of magnitude at $T = 3 \times 10^8$ K and $\varrho = 3000$ gcm$^{-3}$
with respect to the terrestrial value.
Applying the simple expression for the branching strength, it appears
that the branchings are working only at low temperatures (Table~\ref{tab2}),
at least in the straightforward constant-temperature concept of 
the classical model. Since a thermal energy of $kT=29$ keV has instead been 
estimated by that approach \cite{AKW99b,BSA01}, it is obvious that the classical 
model fails in describing the $^{128}$I branching. Indeed, using the new $^{128}$Xe
and $^{130}$Xe cross sections, a fixed temperature of $kT = 29$ keV and 
a fixed matter density of 1.3$\times$10$^3$ g/cm$^{3}$, one obtains an abundance ratio
of ($^{128}$Xe/$^{128}$Xe)$_s$ = 0.51$\pm$0.02, as reported in Table \ref{tab5}. 

This ratio matches the solar ratio, but not the Xe-S ratio of
0.447$\pm$0.003 measured in mainstream presolar grains. This problem can not
be solved, even if one assumes a factor of two for uncertainty of the EC rate of 
$^{128}$I, because this does not change the branching factor by more than 3\%. It is
worthwhile mentioning that in a recent investigation of the uncertainties of the decay 
rates calculated by \cite{TaY87}, the uncertainties of the stellar beta-decay rates 
of $^{128}$I were estimated to be only $\pm$10\% \cite{Gor99}.  

Consequently, it needs to be checked, how this problem can be treated by the 
more comprehensive stellar $s$-process models. 

\section{The $s$ process in thermally pulsing AGB stars \label{models}}

\subsection{Models used}

\begin{figure}
\plotone{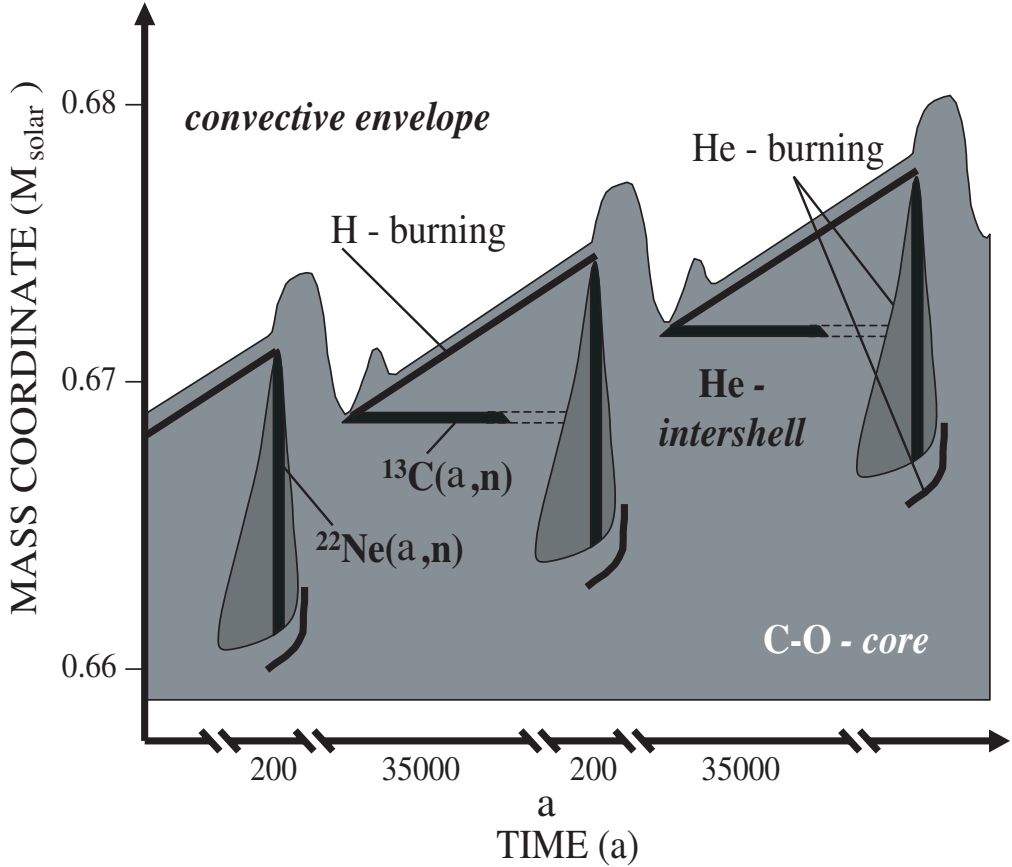}
\caption{Schematic illustration of the $s$-process during the AGB-phase. The breaks 
in the time axis illustrate the shortness of the He shell flashes, lasting only a few 
hundred years, compared with the interpulse phase of about 35,000 yr. The mass
coordinate (in $M_{\odot}$) indicates the thin He intershell, which is the site of the 
$s$ process. The $^{13}$C($\alpha$, n)$^{16}$O reaction represents the dominant neutron 
source, which operates during interpulse period, whereas the higher temperatures 
during the convective He shell flash eventually activate the $^{22}$Ne($\alpha$, n)$^{25}$Mg 
reaction, which is important for establishing the abundance patterns of the $s$-process
branchings \protect\cite{GAB98}.   
\label{fig3}}
\end{figure}

Current stellar models for describing the main $s$-process 
component in the mass range $A \ge$ 90 refer to helium shell burning 
in thermally pulsing low mass AGB stars \cite{IbR83,BGW99}. This scenario 
is characterized by the subsequent operation of two neutron sources 
during a series of helium shell flashes. First, the
$^{13}$C($\alpha$, n)$^{16}$O reaction occurs under radiative conditions
during the intervals between convective He-shell burning episodes \cite{SGB95}. The
$^{13}$C reaction, which operates at low temperatures ($kT \sim$ 8 keV) and 
neutron densities ($n_n \leq 10^7$ cm$^{-3}$), provides most of the 
neutron exposure. The rate of the $^{13}$C reaction has been adopted from 
Denker et al. (1995). However, the resulting abundances are modified
by a second burst of neutrons from the $^{22}$Ne($\alpha$, n)$^{25}$Mg 
reaction, which is marginally activated during the next convective instability,
when higher temperatures ($kT \sim$ 23 keV) are reached in the
bottom layers of the convective pulse, leading to peak neutron densities 
of $n_n \leq 10^{10}$ cm$^{-3}$ \cite{GAB98}. The rate
of the $^{22}$Ne reaction, which determines the strength of
this neutron source, has been adopted from the evaluation of K\"appeler 
et al. (1994), excluding the contribution by the resonance at
633 keV and using the lower limit for the width of the
resonance at 828 keV. Fig.\,\ref{fig3} shows a 
schematic representation of the sequence of He-shell flashes and the 
alternating interpulse periods.

The $^{13}$C is produced via the reaction sequence 
$^{12}$C(p, $\gamma$)$^{13}$N($\beta^+)^{13}$C, when a small amount 
of protons is diffusing from the envelope into the top layers of the He- 
and $^{12}$C-rich zone, forming the so-called $^{13}$C pocket. This 
proton diffusion is supposedly driven by the occurrence of the third 
dredge-up episode, when H burning is temporarily inactive and the convective
envelope penetrates into the upper region of the He intershell.   

Although the second burst accounts only for a few percent of the total 
neutron exposure, it determines the final abundance 
patterns of the $s$-process branchings. This makes the branchings a 
sensitive test for the interplay of the two neutron sources as well as 
for the time dependence of neutron density and temperature during
the second neutron burst. In this context, it is important to note 
that the ($n, \gamma$) cross sections in the Te-I-Xe region are large 
enough that the typical neutron capture times are significantly shorter 
than the duration of the neutron exposure during the He shell flash. 

As far as the isotopic abundance pattern of xenon is concerned, the current set 
of AGB models, covering a range of stellar masses (1.5 $\le M/M_{\odot} \le$ 3) and 
metallicities ($-0.5 \le$ [Fe/H] $\le$ 0), was found to yield 
surprisingly consistent results. Changing the amount of $^{13}$C, and hence
the integrated neutron flux as well as the ($\alpha$,n) rates of $^{13}$C and 
$^{22}$Ne by a factor of two, affected the $^{128}$Xe/$^{130}$Xe ratios by 
less than 1.5\%. This is remarkable, since the $^{13}$C pocket determines the 
actual efficiency for neutron capture nucleosynthesis, whereas 
the $^{22}$Ne rate governs the neutron density during He flashes.
The effects due to variations of the most sensitive nuclear and stellar 
quantities are summarized in Tables \ref{tab3} and \ref{tab4}.

In view of the robustness of the $^{128}$Xe/$^{130}$Xe ratios with
respect to the parameterization of the models used, the current 
calculations of AGB nucleosynthesis were based on the standard  
assumptions, which have been shown to match the solar main 
$s$-process component \cite{AKW99b}, i.e. using
the average of models for 1.5 $M_{\odot}$ and 3 $M_{\odot}$, a 
metallicity of 0.5 $Z_{\odot}$, and the standard $^{13}$C pocket
\cite{BGW99,GAB98}. Keeping the standard parameters unchanged,
the resulting xenon abundances allowed us to study the effect of the 
time scales.

\subsection{Convection in He shell flashes}

Preliminary studies of the convective zone, generated by He shell flashes, 
reported turnover times of a few hours \cite{HoI88}. In this work, 
extensive calculations were carried out in order to study the possible effect on 
the branchings at $^{127}$Te and $^{128}$I more detailed. 

The evolution of the internal structure of AGB stars was calculated for an
initial mass range between 1 and 3 $M_\odot$. Convective velocities are 
evaluated by means of the mixing length theory. The numerical algorithms and 
the input physics of the {\it FRANEC} code used in this work have been extensively 
presented elsewhere \cite{SCL97,CLS98}. 
 
The results for a typical thermal pulse of a 3 $M_{\odot}$ AGB star of solar 
composition are summarized in Fig.\,\ref{fig4}. The temperature in the convective
shell (top) is shown as a function of the mass 
coordinate for two different times (27 days before and 23 days after the maximum 
of the thermal pulse, respectively). The corresponding calculated 
convective velocities are plotted in the panels below, showing the dependence 
on the mass coordinate (middle) and on the internal radius (bottom). The scale 
on the abscissa starts at the bottom of the convective shell. A comparison of the 
latter plots shows how the convective shell expands after the pulse maximum, while
remaining almost constant in mass. The convective turnover time is only 
$\approx$1~hour.  

This picture changes with core mass: the larger the core mass the higher are 
both, the peak of the bottom temperature and the peak of the convective velocity. The 
core mass of about 0.62 $M_{\odot}$ of the model shown in Fig.\,\ref{fig4} is
typical for a low mass AGB star. In fact, during the AGB phase the core mass 
increases from about 0.56 to 0.66 for initial stellar masses ranging from 
1.5 to 3 $M_{\odot}$. Thus, temperatures as well as convective velocities of the 
intershell are increasing during the evolution along the AGB. With increasing 
stellar masses, the core mass at the beginning of the AGB phase 
becomes larger, resulting in a corresponding increase of temperatures and 
convective velocities.

Convective velocities are 
averages, which depend on the degree of super-adiabacity and on the 
adopted mixing length. Since convection is very efficient in stellar interiors,  
the degree of super-adiabacity is very small, leaving the mixing length as the 
main source of uncertainty. As usual, the mixing length parameter 
was calibrated by fitting the solar radius. The mixing length may realistically range 
between one and two pressure scale heights. Accordingly, the convective velocities 
may vary within a factor of two, because they depend linearly on the mixing length.

The short convective turnover time 
ensures that $^{128}$I is efficiently removed from the hot reaction zone
in spite of its 25~min half-life. From Fig. \ref{fig4} one can infer that 
freshly produced material leaves the bottom zone within only 50 to 200~s. 
Accordingly, $^{128}$I decays predominantly at lower temperatures, thus 
favoring the EC branch towards $^{128}$Te, which yields correspondingly 
smaller $^{128}$Xe/$^{130}$Xe ratios.

\section{Results and discussion \label{results}}

The $s$-process nucleosynthesis calculation has been performed using
a post-processing technique that carefully follows the stellar
evolutionary structure up to the tip of the AGB (for details, see
Section 2 of Gallino et al. 1998). 

\subsection{$s$-Process analyses}

The results of the branching 
analyses are summarized in Table \ref{tab5}, which lists the calculated isotopic 
ratios for the $s$-only nuclei $^{128}$Xe and $^{130}$Xe. The respective 
uncertainties are essentially determined by the 1.5\% uncertainty of the cross section ratio 
$\sigma$($^{128}$Xe)/$\sigma$($^{130}$Xe) itself \cite{RHK02} as well as by the 
small uncertainties of the stellar branching ratio of $^{128}$I discussed below.

While the first line of Table \ref{tab5} illustrates the argument  
that the high temperature imposed by the classical model is not compatible 
with the $^{128}$Xe/$^{130}$Xe ratios in Xe-S, 
the last lines refer to the stellar model results, which show a more promising
behavior.
  
Since the decay of $^{128}$I is dominated by the $\beta^-$-mode, the results
are insensitive to the stellar EC rate: Variations by a factor of two affect the 
branching ratio by less than 3\%. This shows the robustness of the stellar model 
against changes in the stellar environment, resulting in an overall uncertainty  
of less than 3\%. 

A first hint that convection in the He shell has a noticeable effect on the
branching was observed by comparing the $^{128}$Xe/$^{130}$Xe ratios obtained 
with the {\it NETZ} code \cite{Jaa91}, which follows the $s$-process network
with the neutron density and temperature profiles from the full AGB model,
but neglects the effect of convection. These results yield $^{128}$Xe/$^{130}$Xe 
ratios that are systematically higher than calculated with the full AGB model.  
  
In the full model, the effect of convection on the abundance evolution was studied 
by reducing the time steps from the usually adopted 10$^6$~s to 10$^5$~s and 
eventually to 3$\times$10$^4$~s (or 8~h). Though this led to a small effect on the 
$^{128}$Te production via the branching at $^{127}$Te, the net effect on the 
final $^{128}$Xe/$^{130}$Xe ratio is negligible due to the rearrangement of
the abundance pattern during freeze-out, when the neutron density is decreasing 
at the end of the pulse. 

The change in the production factors during typical shell flashes 
$$ \frac{^{i}\mbox{Xe}(t) / ^{i}\mbox{Xe}_
\odot}{^{130}\mbox{Xe}_{end} / ^{130}\mbox{Xe}_\odot}$$
are shown in Fig. \ref{fig5} together with the time dependence of the average 
neutron density in the convective zone. The production factors are plotted 
relative to the $^{130}$Xe abundance at the end of the shell flash and are 
normalized to the solar values. Obviously, the increase of temperature at 
the bottom of the convective shell with stellar mass and with pulse number 
is reflected by an increasing peak neutron density released by the 
$^{22}$Ne($\alpha,n$)$^{25}$Mg reaction. 

The variations found in Fig.~\ref{fig5} confirm the more complex situation 
sketched in $\S$4.2 than one might expect from the trends of the branching factor
given in Table \ref{tab2}. During the low temperature phase between He shell 
flashes, the neutron density produced via the $^{13}$C($\alpha$, n)$^{16}$O 
reaction is less than 10$^7$ cm$^{-3}$. The neutron capture branch at 
 $^{127}$Te being completely 
closed results in a $^{128}$Xe/$^{130}$Xe abundance ratio of 0.94 relative
to the solar values at the end of the low temperature phase due to the effect 
of the $^{128}$I branching. After the onset 
of convection and mixing with material from previous flashes at the beginning 
of the He shell flash, one finds the 8\% difference between $^{128}$Xe and 
$^{130}$Xe, which is then modified as shown in Fig.~\ref{fig5}. 

During the following shell flash, the branching remains open
in spite of the higher temperatures of this phase. There are three essential 
effects, which concur to explain this behavior:
\begin{itemize}
\item During the peak of temperature and neutron density, the electron densities at 
the bottom of the convective He shell flash, i.e. in the $s$-process zone, are 
between 15 $\times$ 10$^{26}$ cm$^{-3}$ and 20 $\times$ 10$^{26}$ cm$^{-3}$.
According to Table \ref{tab2}, the branching 
at $^{128}$I is never completely closed. Even at the peak temperatures of the 
He-shell flash, typically 3\% of the flow are bypassing $^{128}$Xe.
\item The quick transport of $^{128}$I from the production zone to cooler layers
implies that the EC decay branch remains significant. This causes more of the reaction 
flow to bypass $^{128}$Xe and leads to smaller $^{128}$Xe/$^{130}$Xe ratios.
\item Around the maximum of the neutron density, the branching at $^{127}$Te is
no longer negligible. Instead, it leads to a significant effect on the 
$^{128}$Xe abundance, which is more pronounced for the 25th pulse of the 3
$M_{\odot}$ model shown in the top panel of Fig.~\ref{fig5}. Correspondingly,
this branching is responsible for an $s$-process contribution of about 3\% to
the abundance of $^{128}$Te, which is considered as an $r$-only isotope.
\end{itemize}

After averaging over the AGB evolution of the two stellar models for
1.5 and 3 $M_{\odot}$ with a metallicity of 0.5$Z_{\odot}$, as 
described by Arlandini et al. (1999), we find an abundance ratio of 
($^{128}$Xe/$^{130}$Xe)$_s$ = 0.466$\pm$0.015. This ratio corresponds to the He
shell material, which is cumulatively mixed with the envelope by the recurrent
third dredge-up episodes and eventually dispersed in the interstellar 
medium by efficient stellar winds. Consequently, it allows
the solar $^{128}$Xe abundance to contain a 9\% $p$-process contribution.
As an estimate for the Galactic average at the formation of the solar system, 
a ratio of ($^{128}$Xe/$^{130}$Xe)$_s$ = 0.468 was obtained by renormalizing 
the result of Travaglio et al. (1999) on the basis of the new Xe(n,$\gamma$) cross sections. 
This value represents the $s$-process evolution in the Galaxy by integrating over all 
previous generations of AGB stars of different masses and
different metallicities. 
The observed ratio of 0.447$\pm$0.003 in the Xe-S component can only be approached, 
if the remaining uncertainties are considered, in particular in the nuclear physics 
data and in the choice of the $^{13}$C pocket (Table \ref{tab4}).

In summary, the combined effects of mass density, temperature, neutron density, 
and convective turnover appear to be consistently described, resulting in the 
successful reproduction of the solar $^{128}$Xe/$^{130}$Xe ratio. This result
represents an additional test of the employed
stellar $s$-process models related to thermally pulsing low mass AGB stars.

Apart from the models discussed, other $s$-process scenarios do not contribute
to the Xe-S problem. In particular, the weak $s$-process component related to 
helium burning in massive stars of 10 to 25 $M_{\odot}$ can be neglected. This was
confirmed by calculations using the {\it NETZ} code with the temperature and neutron
density profiles from a model for a 25 $M_{\odot}$ star \cite{RGB93}. 
Normalization of the resulting abundance distribution to the $s$-only 
nuclei $^{70}$Ge and $^{76}$Se showed that less than 0.4\% of the solar xenon 
abundance could be accounted for in this scenario.  

\subsection{Other processes \label{explosive}}

While the stellar models confirm the isotopic pattern of Xe-S to be, indeed, of 
$s$-process origin, the excess of $^{128}$Xe in solar material has to be 
ascribed to a different source, most likely to the $p$ process. 

In spite of considerable uncertainties, network calculations for 
$p$-process nucleosynthesis in explosively 
burning Ne/O layers of type II supernovae \cite{RAH95} indicate
a significant contribution to the $^{128}$Xe abundance, whereas the $p$ 
production of $^{130}$Xe is much less efficient. By normalizing the 
yields to the $p$-only nuclei $^{124, 126}$Xe, different SN II models 
find $p$~contributions to the solar $^{128}$Xe abundance of 25\% 
(Rayet et al. 1995; M. Rayet, 
private communication) and 12\% \cite{PHR90}, respectively. 
A study of the $p$ process in type Ia supernovae 
\cite{HMW91} yielded a $p$-abundance ratio $^{124}$Xe/$^{126}$Xe that
differs by a factor of three from the solar value, corresponding to an 8\% 
$p$ contribution to the solar $^{128}$Xe abundance. Recently, full 
nucleosynthesis calculations following the hydrostatic and explosive 
phases of massive stars by Rauscher et al. (2002) have provided 
similar results for an enhanced $p$-process yield of $^{128}$Xe 
compared to $^{130}$Xe.

Contributions from alternative nucleosynthesis mechanisms,  
like the $rp$ and the $\nu$-processes, can be excluded. The $rp$~process can
be ruled out, because reaction path ends at $A\approx107$ in the Sn-Sb-Te cycle \cite{SAB01}.
Detailed calculations showed that the $\nu$ process 
contribution to the xenon abundance distribution is negligible \cite{WHH90}. 
 
\section{Summary  \label{conclusions}}

The solar abundances of $^{128}$Xe and
$^{130}$Xe are produced by the main component of the $s$ process, except for a
non-negligible $p$ contribution to $^{128}$Xe. By comparison with pure $s$-process
xenon, i.e. the Xe-S found in presolar grains, this $p$ component amounts to
9$\pm$3\%, thus providing an additional constraint for $p$-process calculations.
 
The $s$-process reaction flow through the branchings at $^{127}$Te and 
$^{128}$I has been followed in detail with stellar evolutionary models 
for the asymptotic giant branch (AGB) phase. It was found that the solar abundances 
as well as the Xe-S ratio of $^{128}$Xe and $^{130}$Xe could be successfully 
reproduced. Since these branchings exhibit a much weaker dependence  
on neutron density than other cases, they represent an important complement
to all previous branching analyses \cite{AKW99b,BSA01}. Apart from an small additional 
effect of the electron or mass density it turned out that the short half-life 
of the $^{128}$I branch point represents a first constraint for the convective 
velocities during the He shell flash. In view of the complex interplay 
of all these parameters, the consistent description of the $^{128}$Xe/$^{130}$Xe 
ratio can be considered as a further successful test of the AGB models used. 

The lower $^{128}$Xe/$^{130}$Xe ratio measured in presolar mainstream SiC 
grains compared to the solar abundance ratio corresponds to the situation in 
thermal pulses during the last part of the AGB phase, which exhibit higher 
temperatures and neutron densities. This confirms that the mainstream SIC grains, 
the carriers of the Xe-S meteoritic component, were predominantly formed 
during this epoch.

The impact of the fast convective turnover time scale found in the present
study is also of potential interest for other, previously neglected branchings.

\acknowledgments

We would like to thank U. Ott, R. Pepin, and M. Rayet for clarifying discussions.
R.R. is indebted to CERN for support by the doctoral student program. 
This work was partly supported by the Italian MIUR-FIRB grant "The astrophysical 
origin of the heavy elements beyond Fe". 

\newcommand{\noopsort}[1]{} \newcommand{\printfirst}[2]{#1}
  \newcommand{\singleletter}[1]{#1} \newcommand{\swithchargs}[2]{#2#1}


\begin{deluxetable}{cccc}
\tabletypesize{\scriptsize}
\tablecaption{Different abundance ratios of $^{128}$Xe and 
              $^{130}$Xe          \label{tab1}}
\tablewidth{0pt}
\tablehead{
		   \colhead{Abundance ratio $^{128}$Xe/$^{130}$Xe} & 
		   \colhead{Reference}   &
		   \colhead{Remark}
		  }
\startdata

0.466 $\pm$ 0.014		   &  \cite{DeB91}  & terrestrial     \\
0.503	 				   &  \cite{AnG89}  & solar 		  \\
0.510 $\pm$ 0.005		   &  \cite{PBR95}  & solar 		  \\
0.447 $\pm$ 0.003		   &  \cite{LAA94}  & Xe-S (SiC) 		  \\
0.586 $\pm$ 0.004		   &  \cite{HuL94}  & Xe-HL (Diamond) 	  \\
\enddata
\end{deluxetable}

\begin{deluxetable}{ccccccc}
\tabletypesize{\scriptsize}
\tablecaption{Beta-decay branching ratio at $^{128}$I as a function of electron
              density and temperature \cite{TaY87}. \label{tab2}}
\tablewidth{0pt}
\tablehead{
		   \colhead{Electron density ($10^{26}$ cm$^{-3}$)} &
		   \multicolumn{6}{c}{Temperature (10$^8$ K)}
		  }
\startdata

   & 0     & 1     & 2     & 3     & 4     & 5 		\\
0  & 0.940 & 0.963 & 0.996 & 0.999 & 1.000 & 1.000 	\\
3  & 0.940 & 0.952 & 0.991 & 0.997 & 0.999 & 0.999 	\\
10 & 0.940 & 0.944 & 0.976 & 0.992 & 0.996 & 0.997 	\\
30 & 0.940 & 0.938 & 0.956 & 0.980 & 0.989 & 0.995  \\
\enddata
\end{deluxetable}

\begin{deluxetable}{cccccccc}
\tabletypesize{\scriptsize}
\tablecaption{The $s$-process $^{128}$Xe/$^{130}$Xe ratios obtained by variation
         of the most sensitive nuclear quantities \label{tab3}}
\tablewidth{0pt}
\tablehead{
\multicolumn{2}{c}{$^{128}$I EC-rate$^a$} \quad	& \quad & 
\multicolumn{2}{c}{$^{13}$C($\alpha$,n)$^{16}$O rate$^b$}     \quad & \quad & 
\multicolumn{2}{c}{$^{22}$Ne($\alpha$,n)$^{25}$Mg rate$^c$}\\
\cline{1-2}  \cline{4-5} \cline{7-8}\\

[Fe/H] = $-$ 0.30 & ($^{128}$Xe/$^{130}$Xe)$_{s}$ \quad & \quad & \quad
[Fe/H] = $-$ 0.30 & ($^{128}$Xe/$^{130}$Xe)$_{s}$ \quad & \quad & \quad
[Fe/H] = $-$ 0.30 & ($^{128}$Xe/$^{130}$Xe)$_{s}$ }
 
\startdata

SR		& 0.466	\quad & \quad & \quad SR	& 0.466	\quad & \quad & \quad SR   & 0.466 \\
SR/2    & 0.482	\quad & \quad & \quad SR*2	& 0.471 \quad & \quad & \quad SR*2 & 0.470 \\
--		& -- 	\quad & \quad & \quad SR/2	& 0.468	\quad & \quad & \quad SR/2 & 0.466 \\
\enddata

$^a$ standard rate (SR) from Takahashi \& Yokoi (1987) \\
$^b$ standard rate (SR) adopted from Denker et al. (1994)\\
$^c$ standard rate (SR) adopted from K\"appeler et al. (1994) as described in text\\
\end{deluxetable}

\begin{deluxetable}{ccccc}
\tabletypesize{\scriptsize}
\tablecaption{The $s$-process $^{128}$Xe/$^{130}$Xe ratios obtained by variation
         of the most sensitive stellar quantities \label{tab4}}
\tablewidth{0pt}
\tablehead{
\multicolumn{2}{c}{$^{13}$C-pocket$^a$} \quad & \quad & 
\multicolumn{2}{c}{Metallicity}      \\
\cline{1-2}  \cline{4-5} \\

[Fe/H] = $-$ 0.30 & ($^{128}$Xe/$^{130}$Xe)$_{s}$ \quad & \quad & \quad
[Fe/H]	          & ($^{128}$Xe/$^{130}$Xe)$_{s}$ 
 }
 
\startdata

ST*2		& 0.446	\quad & \quad & \quad 0.0		& 0.460	\\
ST		    & 0.466	\quad & \quad & \quad $-$ 0.30	& 0.466 \\
ST/2		& 0.465	\quad & \quad & \quad $-$ 0.52	& 0.461	\\
\enddata

$^a$ standard case (ST) chosen to match overall solar $s$ abundances \cite{GAB98}	\\
\end{deluxetable}

\clearpage

\begin{deluxetable}{lcl}
\tabletypesize{\scriptsize}
\tablecaption{Results of the $\protect s$-process calculations \label{tab5}}
\tablewidth{0pt}
\tablehead{
                                    \colhead{Model} 
		                          & \colhead{Abundance ratio $^{128}$Xe/$^{130}$Xe} 
		                          & \colhead{Remarks}}
\startdata
                                     
classical model       & 0.51 $\pm$ 0.02   & $kT=29$ keV, $^{127}$Te thermalized,\\
				      &                   & {\it NETZ} code$^a$        \\
simplified AGB        & 0.477 $\pm$ 0.02   &  turnover neglected, {\it NETZ} code$^a$ \\
AGB$^b$, full network     & 0.473 $\pm$ 0.015 & time steps of 10$^6$ s     \\
AGB$^b$, full network     & 0.466 $\pm$ 0.015 &  time steps of 10$^5$ s    \\
\enddata

$^a$ \cite{Jaa91}\\
$^b$\cite{GAB98}\\
\end{deluxetable}
\clearpage

\begin{figure}
\plotone{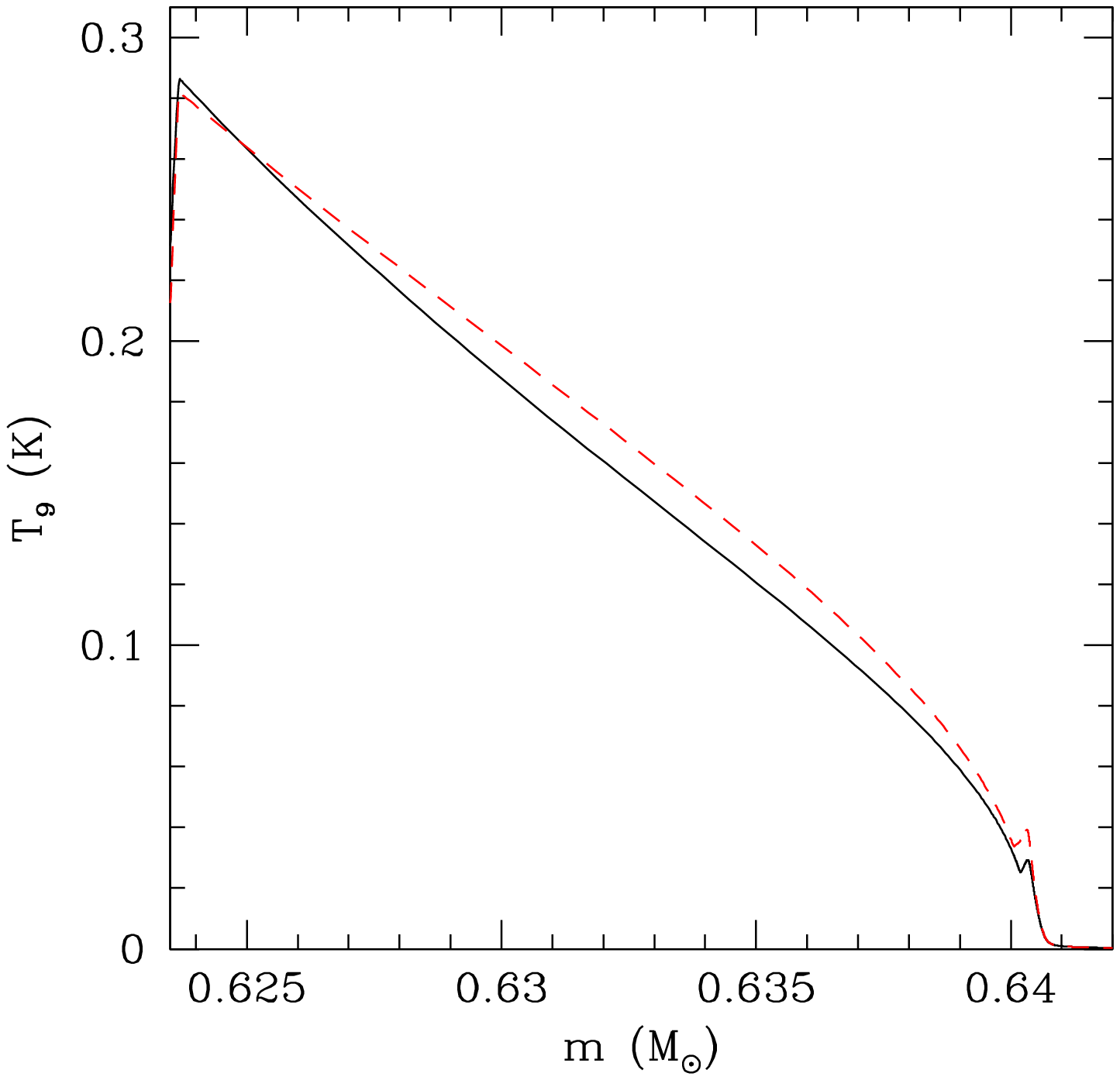}
\end{figure}

\begin{figure}
\plotone{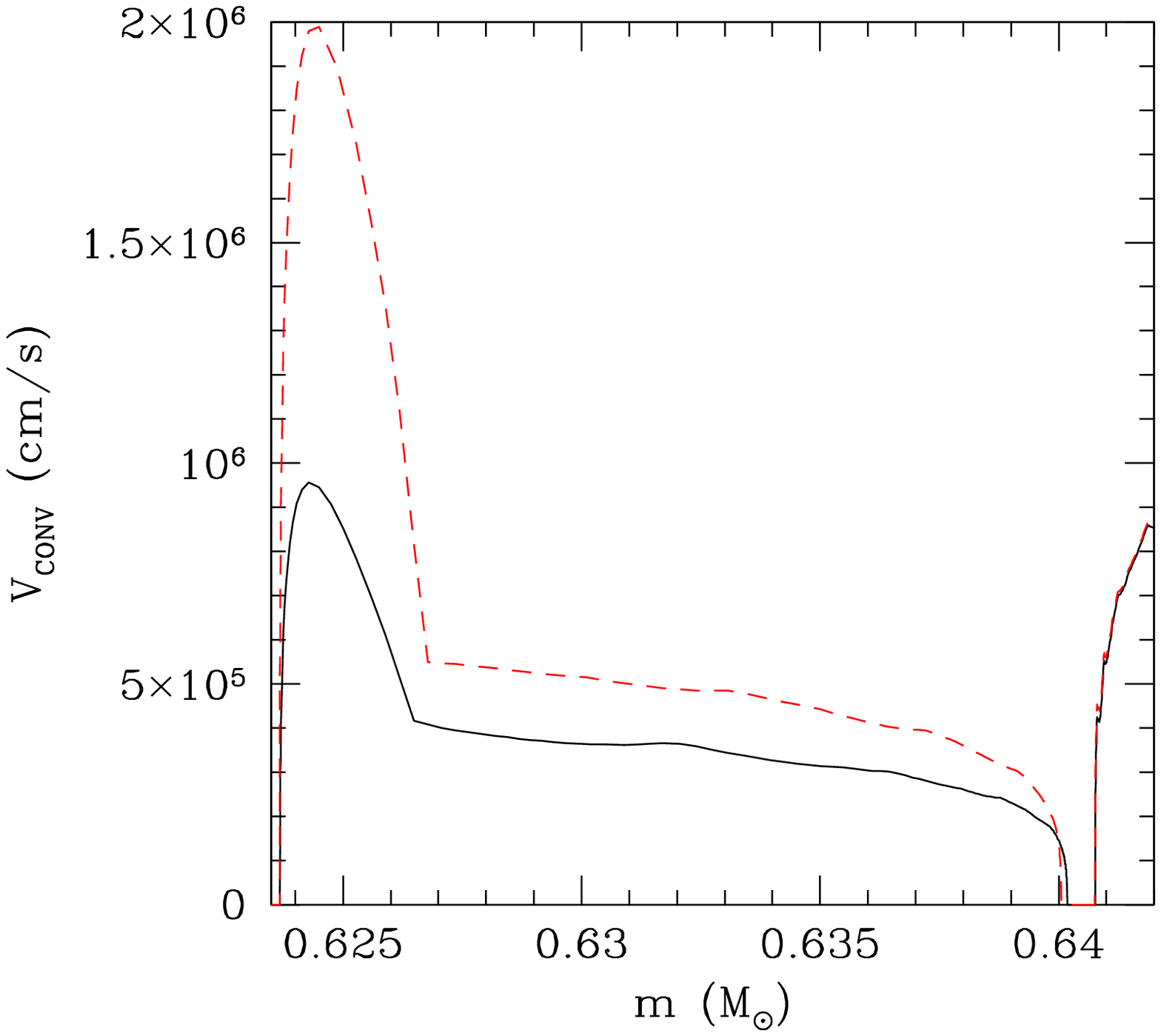}
\end{figure}

\begin{figure}
\plotone{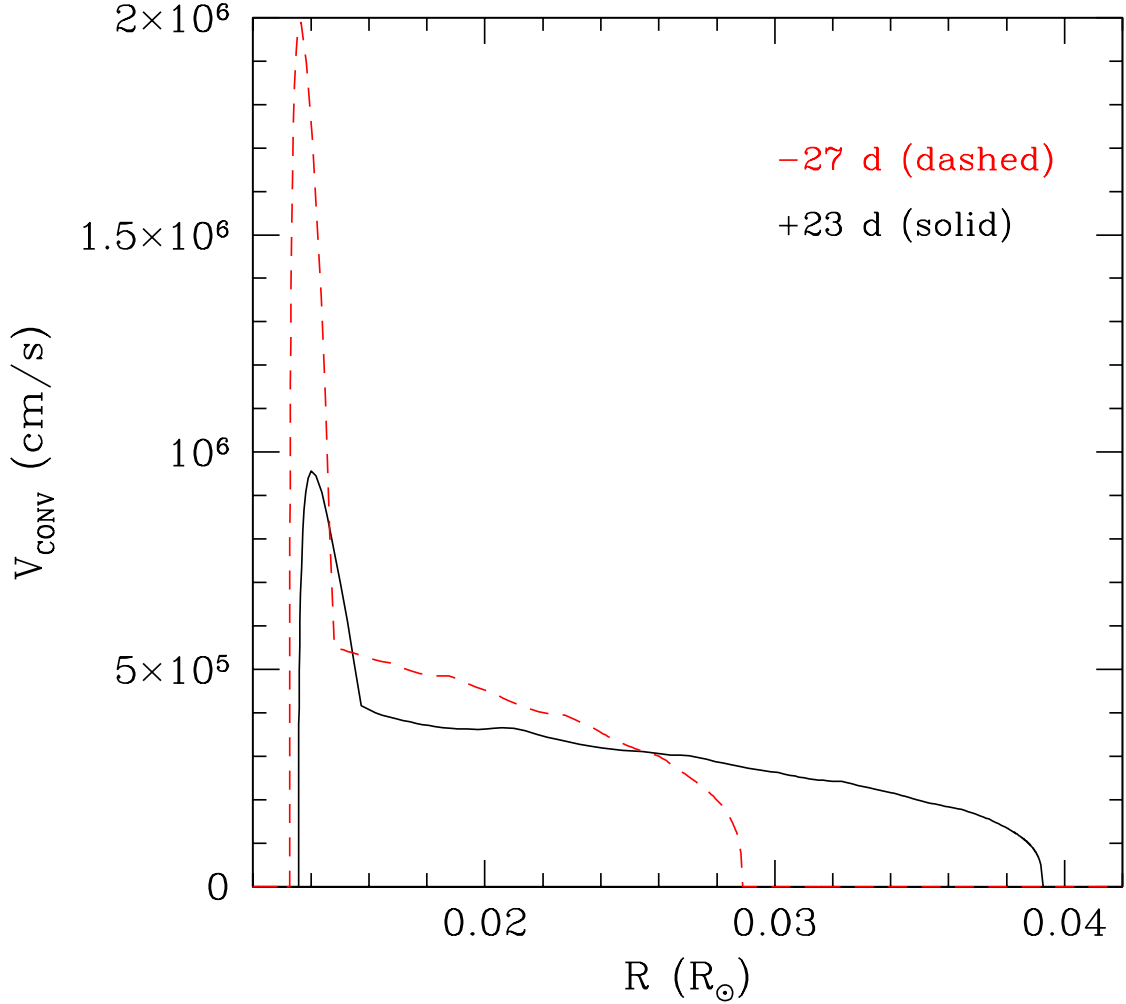}
\caption{
The temperature in the convective shell in a 3 $M_\odot$ star of solar
composition during a typical thermal pulse (top) as a function of the mass 
coordinate for two different times, 27 d before and 23 d after pulse maximum  
(dashed and solid lines, respectively). The abscissa 
starts at the bottom of the convective shell. The corresponding 
convective velocities as a function of the mass coordinate (middle) and of 
the internal radius (bottom) are plotted below. The convective turnover 
time is $\approx$1 hour, but it takes only 50 to 200 s to transport freshly 
produced $^{128}$I from the $s$-process zone in the bottom layers into the 
outer, cooler regions.
\label{fig4}}
\end{figure}

\begin{figure}
\plotone{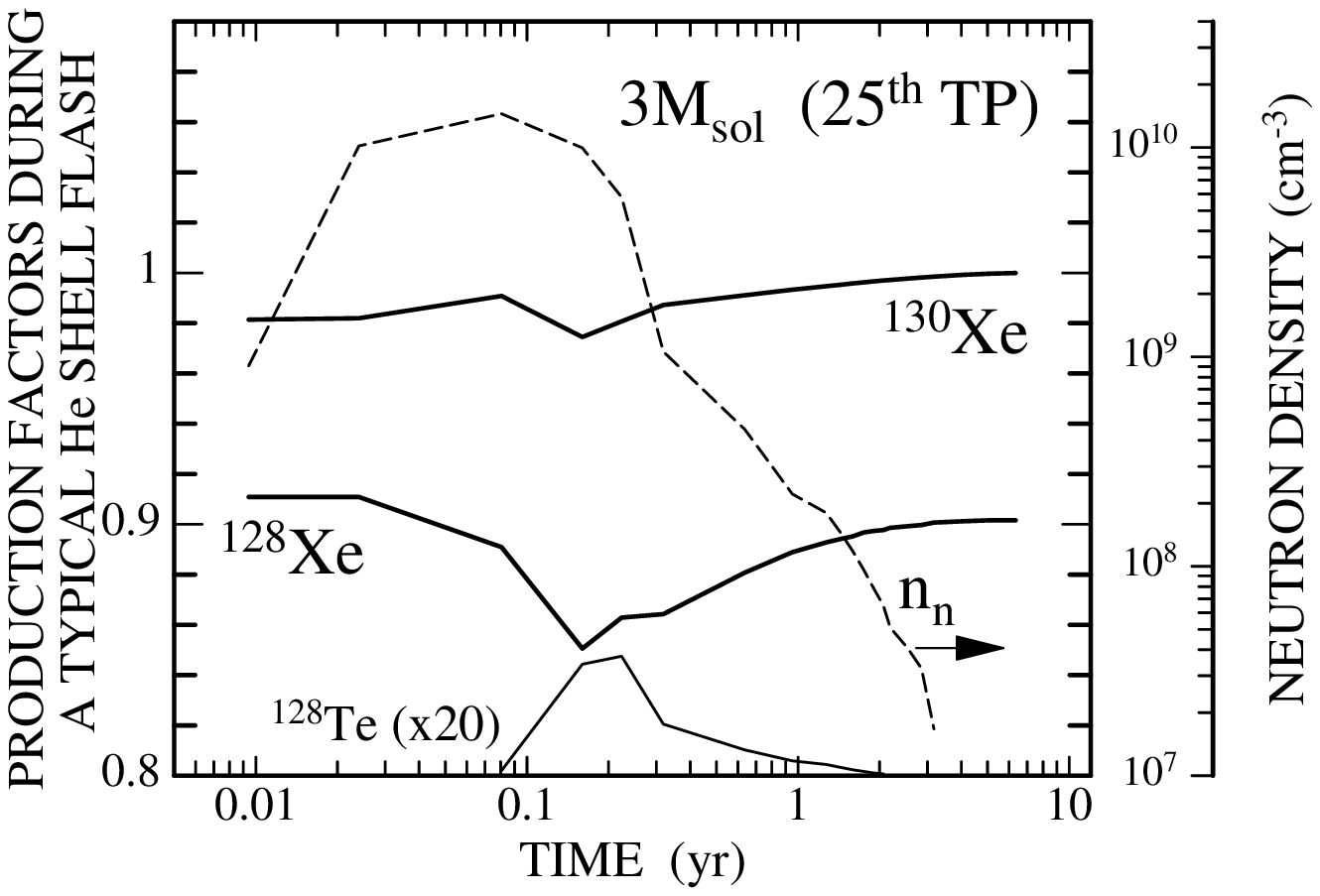}
\end{figure}

\begin{figure}
\plotone{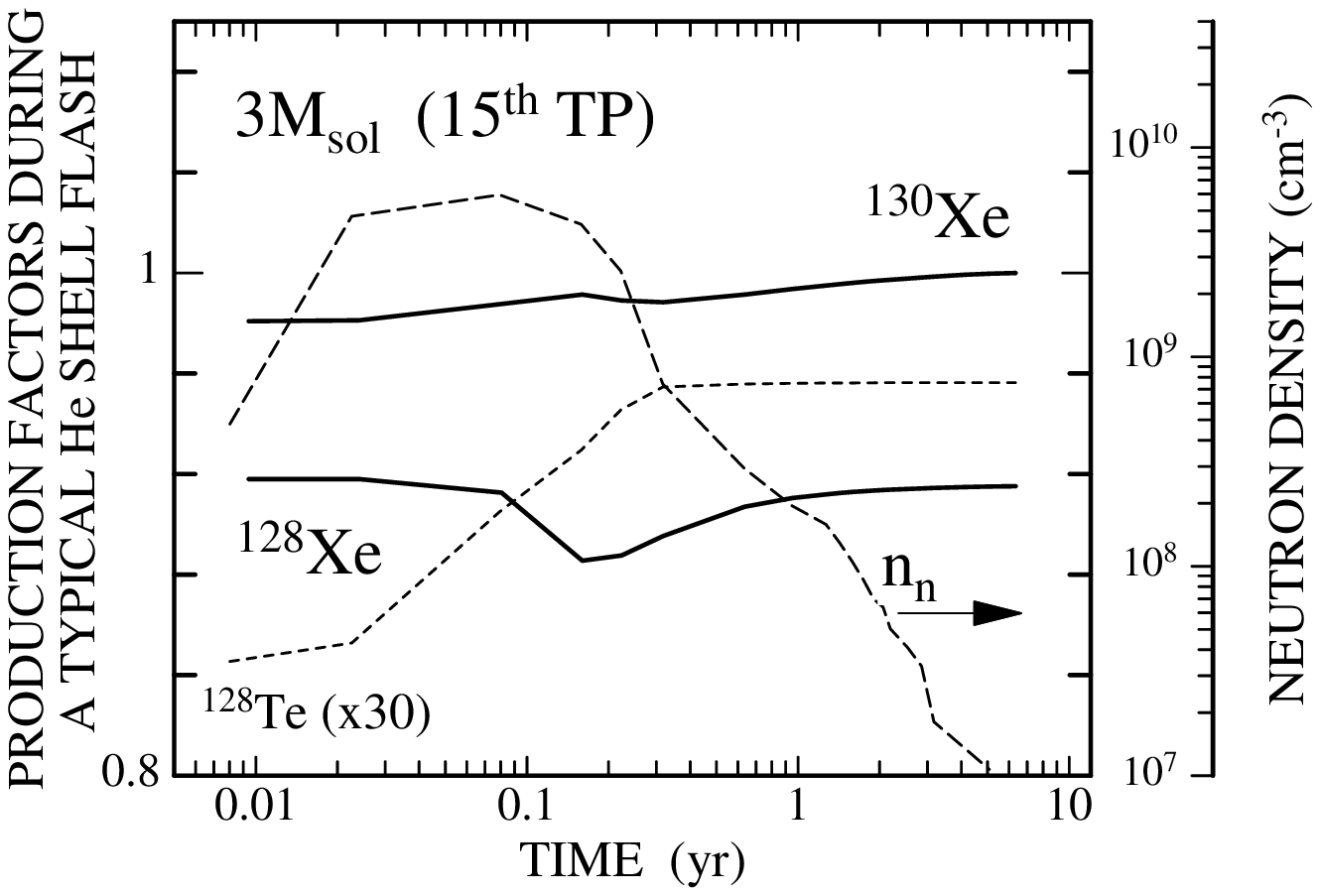}
\end{figure}

\begin{figure}
\plotone{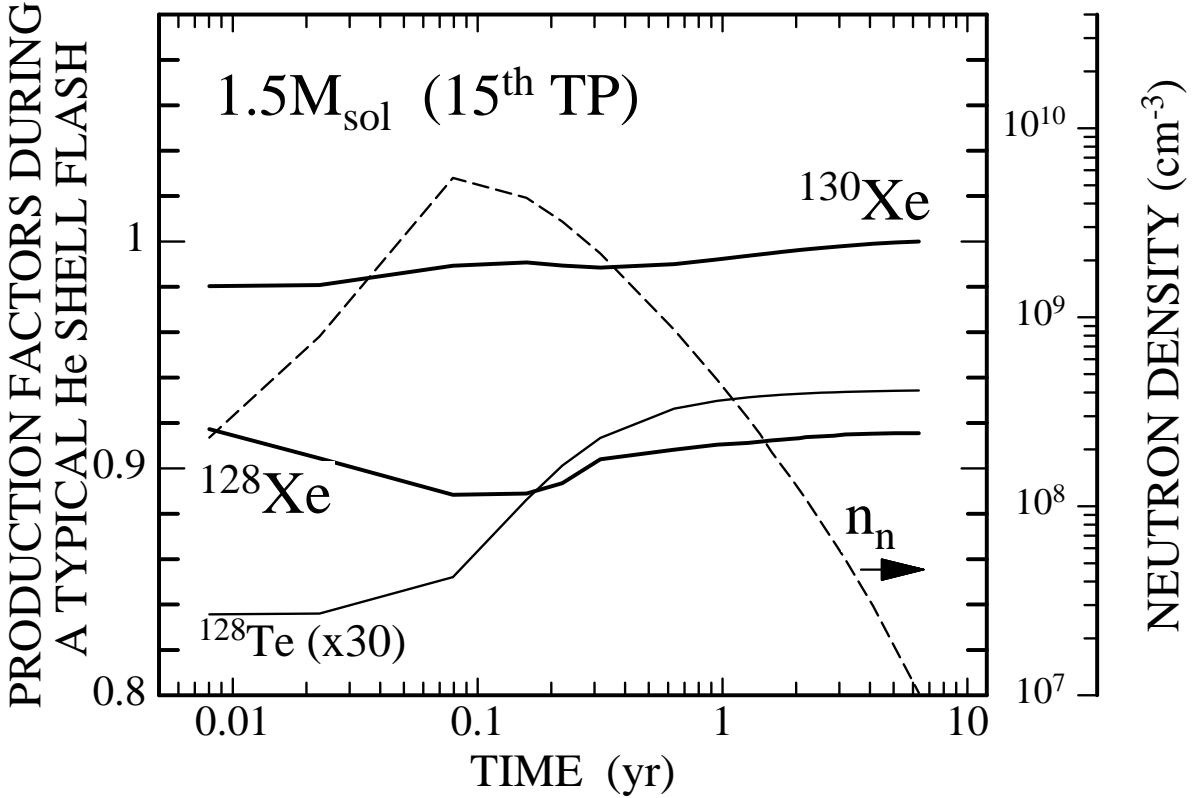}
\caption{The production factors of $^{128}$Xe and $^{130}$Xe during typical
He shell flashes in two AGB stars (top and middle: 25th and 15th pulse in a 3 
$M_{\odot}$ star, bottom: 15th pulse in a 1.5 $M_{\odot}$ star). The time
scale starts when the temperature at the bottom of the convective shell 
reaches 2.5$\times$10$^8$ K, i.e. at the onset of the $^{22}$Ne($\alpha$, 
n)$^{25}$Mg reaction. The curves are normalized to $^{130}$Xe at the end 
of the shell flash as explained in the text. The average neutron density 
produced at the bottom of the convective region 
by the $^{22}$Ne($\alpha$, n)$^{25}$Mg reaction is indicated by the dashed 
line. The production ratio of $^{128}$Xe/$^{130}$Xe is always below unity,
which implies that the branchings 
at $A$=127/128 remain active throughout the shell flash due to the combined 
effects of mass density, temperature, neutron density, and convective turnover 
times (see text). 
\label{fig5}}
\end{figure}

\end{document}